\documentclass{article}
\usepackage[a4paper, total={7in, 10in}]{geometry}
\usepackage{graphicx, subcaption}
\usepackage{amsmath, amssymb, amsthm, bm, xcolor}
\usepackage[backend=biber, style=apa]{biblatex}
\DeclareLanguageMapping{american}{american-apa}
\title{When Teachers Smile or Frown: A Profile-Based Analysis of
Achievement Emotions}
\author{Rudra Mukhopadhyay, Satyaki Mazumder, Koel Das}
\date{January 2026}

\begin{document}

\maketitle

\begin{abstract}
    Achievement emotions shape how students engage with and learn from academic tasks, yet most studies examine individual emotions rather than co-occurring affective profiles and their dynamics. We examined latent achievement-emotion profiles and their transitions following exposure to different instructor facial expressions during a video lecture. Self-reported data from 78 Grade VII and VIII students revealed three profiles: enthusiastic, demotivated, and vulnerable. Profile transitions differed across instructor conditions, with happy expressions favouring more adaptive transitions and angry expressions favouring transitions toward demotivation. Exploratory factor analysis and Bayesian structural modelling further identified preparedness and cognitive restraint as regulatory dimensions associated with profile switching.
\end{abstract}

\section{Introduction}

Learning is a cognitive process that is strongly influenced by emotion. Emotions such as enjoyment, anxiety, boredom, and curiosity shape how students engage with academic tasks and how they learn from them.\\

\noindent
\cite{Pekrun2006, Pekrun2024} defined achievement emotions as emotions directly related to learning activities and their outcomes, including those experienced while learning as well as those arising in anticipation of or in response to academic success or failure. The Control-Value Theory (CVT) of achievement emotions proposes that achievement emotions such as enjoyment, pride, anxiety, shame, boredom, and hopelessness arise from learners' appraisals of control (e.g., perceived competence) and value (e.g., intrinsic motivation) in academic contexts \parencite{Pekrun2006, Pekrun2024}. \cite{Pekrun2024} classified these emotions primarily by valence (positive or negative) and arousal (activating or deactivating). For instance, enjoyment is a positive activating emotion, whereas boredom is a negative deactivating emotion \parencite{Pekrun2024}. CVT further proposes that these emotions are embedded within motivational and cognitive processes: they influence learners' motivation, self-regulation, learning strategies, and academic performance, while also being shaped by students' perceptions of control and value \parencite{Pekrun2024, Pekrun2022_development, Iqbal2023}. Meta-analytic evidence from \cite{Qi2025_recent_meta, CamachoMorles2021_recent_meta} establishes an association between positive activating emotions (e.g., enjoyment, hope, and pride) and better learning performance, whereas negative emotions (e.g., boredom, anger, and hopelessness) are generally linked to poorer outcomes. Positive emotions tend to foster intrinsic motivation and deeper learning strategies, whereas negative emotions can reduce motivation and encourage superficial processing \parencite{Pekrun2022_development, Qi2025_recent_meta}. However, these effects are not absolute, as excessive positive emotions or moderate levels of certain negative emotions (e.g., anxiety) can sometimes hinder or facilitate learning depending on the context \parencite{Pekrun2024, Wu2022, Liu2021}.\\

\noindent
A major part of students' learning occurs inside classrooms. These are inherently social environments, where teachers' verbal as well as non-verbal expressions and behaviours act as powerful signals that can shape students' affective experiences \parencite{Kesevan2020, Liu2021}. Still, facial expressions have remained an underexplored aspect of school instruction and are often considered to have only a marginal influence. \cite{Marici2025} showed that earlier studies have likewise suggested that non-verbal cues produce relatively weak effects. This concern is particularly relevant in computer-based learning environments, where direct interaction is limited and subtle cues such as teachers' facial expressions may play a more important role in shaping students' experiences. Evidence \parencite{Wang2022important, Gu2024, Polat2022} suggests that instructors' expressiveness can influence students' engagement, social presence, and longer-term retention, even when immediate performance effects are weak or absent. For instance, \cite{Marici2025} shows that teachers' positive facial expressions have been associated with greater student engagement, compliance, and more favourable perceptions of the teacher, whereas anger and emotional neutrality generally reduce participation and interpersonal warmth. Experimental work in video learning likewise suggests that expressive instructors can alter students' social presence, arousal, cognitive load, and subsequent learning \parencite{Wang2022important}. However, most studies have focused on isolated outcomes, and no clear consensus exists regarding the affective changes elicited by teachers' displayed emotions \parencite{Mega2014, Wu2022, Zhang2023}.\\

\noindent
In recent years, \cite{Pekrun2006, Robinson2020_profscience, Radii2024_profprimary} have shown that students rarely experience emotions in isolation. Instead, multiple emotions often co-occur, forming distinct emotional profiles that provide a more realistic representation of learners' mental states than approaches examining individual emotions separately. \cite{Pekrun2006} proposed that such mixed emotional experiences commonly arise when learners are uncertain about their likelihood of success, and \cite{Robinson2020_profscience, Robinson2017_profcollege} have confirmed that students may simultaneously experience positive and negative emotions while performing challenging academic tasks. These co-occurring emotions can produce ``blunted or qualitatively altered'' outcomes \parencite{Robinson2017_profcollege}. Although mixed emotional experiences have become an important area of achievement emotion research, most studies have focused on static emotional profiles or mean-level changes, leaving the dynamics of transitions between affective profiles during learning relatively unexplored \parencite{Tze2020_profileuni, Robinson2020_profscience}.\\

\noindent
To address this gap, a small subset of studies focusing specifically on achievement emotions has employed cluster-based analysis, including k-means, Ward's method, Gaussian Mixture Models and their derivatives (e.g., Latent Profile Analysis and Latent Class Analysis). These methods are increasingly preferred because they capture unobserved data heterogeneity \parencite{Hickendorff2018}, account for the co-occurrence of emotions \parencite{Robinson2020_profscience, Tze2020_profileuni, Radii2024_profprimary}, and can accommodate complex multidimensional datasets \parencite{Wang2021_profefl}. These studies \parencite{sachisthal2021trait, tempelaar2022types, Radii2024_profprimary} have identified between two and seven achievement emotion profiles, although most studies commonly report three or four clusters. Across diverse age groups, subjects, and educational contexts, clustering approaches consistently identify positive, negative, and mixed emotional profiles \parencite{Chen2025_profmath, Symes2025_profprimary, Jarrell2016, Robinson2017_profcollege, Tze2020_profileuni}. Studies \parencite{Robinson2020_profscience, Radii2024_profprimary} have further reported context-specific mixed profiles, where positive and negative emotions coexist, such as `Apprehensive-Happy'', `Moderate-High All'', and ``Deactivated'' (pleasantly happy but bored), highlighting the prevalence of mixed emotional experiences.\\

\noindent
However, identifying emotional profiles alone does not explain why students move between them during learning. One important possibility is that instructional experiences affect achievement emotions indirectly by altering the motivational and cognitive conditions under which learning occurs. This interpretation follows directly from CVT, which positions perceived control and value as proximal antecedents of achievement emotions and links them to subsequent motivation, information processing, and achievement \parencite{Pekrun2006, Pekrun2024}. In particular, students who feel capable of succeeding in a task are more likely to experience adaptive emotions such as enjoyment and less likely to experience anxiety, boredom, or hopelessness. Likewise, perceiving a task as useful or personally valuable can strengthen adaptive engagement, while simultaneously making failure more consequential and thereby amplifying some negative emotions under conditions of low control. Empirical studies support this general pattern: higher control and value appraisals have been associated with greater enjoyment and lower boredom or anxiety, as well as with better academic outcomes \parencite{Pekrun2006, Liu2021}. Thus, variables reflecting students' preparedness for a learning task-particularly confidence in their ability and perceived utility of the material-may represent an important motivational context in which achievement emotions are formed and subsequently altered.\\

\noindent
This regulatory role is not confined to motivational appraisal. Achievement emotions are closely linked to students' engagement and the cognitive demands imposed by learning. Positive emotions such as enjoyment can promote persistence, attention, and deeper engagement, whereas boredom and anxiety can undermine sustained engagement and interfere with the allocation of cognitive resources \parencite{Mega2014, Pekrun2022_development, Robinson2017_profcollege}. Conversely, the cognitive demands of learning can themselves shape emotional experience. Cognitive Load Theory distinguishes intrinsic load arising from task complexity from extraneous load arising from unnecessary demands imposed by the instructional environment \parencite{sweller2019}. Studies of achievement emotions suggest that enjoyment is associated with more adaptive patterns of cognitive engagement, whereas boredom and anxiety can accompany or exacerbate less efficient processing \parencite{Sugiyo2018, Sugiyo2021}. In this sense, motivational and cognitive processes are not independent of emotion; rather, they form a reciprocal system in which emotional states influence engagement and processing while the experienced demands of learning can feed back into subsequent emotional states.\\

\noindent
The social context of learning provides another potential regulatory pathway, particularly in computer- and video-based instruction. Social presence refers to the extent to which an instructor or learning partner is perceived as a real person in mediated communication \parencite{gunawardena1995}. In video learning, instructor visibility and expressive non-verbal cues can strengthen social presence, increase attention and arousal, and reduce the sense of interpersonal distance that characterizes mediated instruction \parencite{Wang2022important, Polat2022, Suen2024}. Experimental evidence further suggests that expressive instructor behaviour can increase perceived social presence while altering cognitive load and improving longer-term learning \parencite{Wang2022important}. Social presence may therefore provide a mechanism through which an instructor's facial expression becomes psychologically meaningful to the learner rather than remaining a purely perceptual feature of the video.\\

\noindent
Taken together, these findings suggest that transitions between achievement-emotion profiles may reflect more than a direct response to an instructor's displayed emotion. Instructor expressions may instead alter students' motivational, cognitive, and social experience of the learning episode, which in turn may influence the configuration of achievement emotions. From this perspective, students' confidence and perceived utility can be understood as indicators of their preparedness and control-value appraisals entering the learning episode, whereas motivation, engagement, cognitive load, attention, and social presence capture aspects of the post-lecture cognitive-affective state through which instructional experiences may be translated into emotional change. Such a framework also provides a natural bridge between profile-based analysis and latent-variable modelling: rather than treating achievement emotions and regulatory variables as isolated outcomes, their relations can be examined as components of a common system.\\

\noindent
The present exploratory study therefore examines achievement emotions at two complementary levels. First, we investigate whether students' self-reported emotions form coherent and interpretable affective profiles, and whether these profiles resemble configurations previously reported across different educational populations. Second, we examine whether these profiles change following exposure to different instructor facial expressions in a short video lecture. Finally, we investigate potential regulatory mechanisms underlying these transitions by examining how students' preparedness and subject confidence, perceived utility, and post-lecture cognitive-affective states are organized into latent dimensions and how these dimensions relate to profile switching.\\

\noindent
Specifically, the study addresses the following research questions:

\begin{enumerate}
\item Whether students' self-reported achievement emotions give rise to stable and interpretable profiles, and how these compare with existing findings in the literature.
\item How instructor facial expressions influence transitions between adaptive and maladaptive emotional profiles following a short video lecture.
\item Whether preparedness-related appraisals (general confidence, topic-specific confidence, and perceived utility) and post-lecture cognitive-affective states (motivation, engagement, cognitive load, attention, and social presence) form distinct latent dimensions.
\item Whether these latent regulatory dimensions are associated with emotional profile switching and therefore provide a plausible mechanism linking instructor emotional expression to changes in students' achievement-emotion profiles.
\end{enumerate}

\section{Materials and Methods}
\subsection{Participants}
Eighty participants ($N_0 = 80$) from Grades VII and VIII were recruited from a single CBSE-affiliated school in West Bengal, India. All participants had normal or corrected-to-normal vision, no diagnosed learning disorders, and were proficient in speaking, reading, and writing in English. Written informed consent was obtained from their legal guardians prior to participation. The study procedures were reviewed and approved by the Institutional Ethics Committee. 
Two participants were excluded due to incomplete responses. The final sample comprised seventy-eight participants ($N = 78$).

\subsection{Experimental Paradigm}
The instructional content selected for the video lecture was ``Cell: its organelles and functions". Three versions of the lecture were recorded, in which the instructor displayed distinct facial expressions corresponding to joyous, angry, and neutral emotional states, while the conceptual content, structure, and duration were held constant. Each video was approximately six to seven minutes in length. To enhance ecological validity and simulate a classroom environment, the videos incorporated audio tracks of student responses and the instructor's corresponding exclamatory remarks. 
The experiment was conducted in the school's computer laboratory and comprised four primary components: (i) pre- and post-experiment handwritten subject tests, (ii) pre- and post-lecture assessments of five achievement emotions (joy, interest, hopelessness, boredom, and anxiety), (iii) questionnaires assessing perceived preparedness—measured through general confidence, topic-specific confidence, and perceived utility—and selected affective-cognitive state factors, including motivation, engagement, perceived social presence, and internal and external cognitive load, and (iv) the video lecture intervention.

\subsection{Procedure}
Participants were first briefed about the experimental procedure and given five to ten minutes to ask questions. They then completed a pre-test assessing both factual and comprehension-based knowledge of ``Cell: its organelles and functions". The difficulty level of the test items was evaluated by three subject experts from the institute’s Department of Biological Sciences. 

Following the pre-test, participants were seated individually at a computer and provided self-reported ratings of their current achievement emotions and perceived preparedness using a single-item computer-based questionnaire derived from the shortened version of the Achievement Emotion Questionnaire (AEQ-S) \parencite{Bieleke2021}.

Participants then watched the pre-recorded video lecture on the same topic. The videos were preloaded such that each participant viewed one version featuring either a joyous, angry, or neutral instructor. 
After the lecture, participants again reported their perceived achievement emotions and rated selected affective-cognitive factors, as illustrated earlier. These affective-cognitive variables were measured by a single-item computer-based questionnaire derived from the study by Wang et al. \cite{Wang2022important}.

Finally, participants completed a post-test of comparable difficulty to the pre-test, verified by the same three subject experts mentioned earlier.

\subsection{Behavioural Data Collection and Preliminary Analysis}
The achievement emotion questionnaires were computer-based, using clickable multiple-choice items on a 7-point Likert scale. Participants completed a brief practice session before the experiment to familiarize themselves with the interface. Preliminary analyses on the self-reported emotional ratings and the test scores involved Wilcoxon signed-rank (for pre-post comparison) and Kruskal-Wallis test (for independent sample comparisons). For the Gaussian mixture model (GMM) analyses, the data were mean-centred and range-scaled.

\subsection{Gaussian Mixture Model-based Clustering}
Let $\mathbf{x} \in \mathbb{R}^{d}$ be a $d$-dimensional observation. A \emph{Gaussian Mixture Model} (GMM) with $K$ components assert that each observation is drawn from a convex combination of $K$ multivariate Gaussian distributions. In other words, the observation vector $\mathbf{x}$ can be assumed to be generated by a ``mixture" of many Gaussian distributions with varying means and covariance matrices.
\\
A random vector $\mathbf{x}$ follows a $K$-component GMM if its probability density function is
\begin{equation}
    p(\mathbf{x} \mid \bm{\theta}) = \sum_{k=1}^{K} \pi_k \, \mathcal{N}\!\left(\mathbf{x} \mid \bm{\mu}_k, \bm{\Sigma}_k\right),
    \label{eq:gmm}
\end{equation}
where $\mathcal{N}(\mathbf{x} \mid \bm{\mu}_k, \bm{\Sigma}_k)$ denotes the multivariate Gaussian density
\begin{equation}
    \mathcal{N}\!\left(\mathbf{x} \mid \bm{\mu}_k, \bm{\Sigma}_k\right)
    = \frac{1}{(2\pi)^{d/2} \lvert\bm{\Sigma}_k\rvert^{1/2}}
      \exp\!\left(-\tfrac{1}{2}(\mathbf{x}-\bm{\mu}_k)^{\top}
      \bm{\Sigma}_k^{-1}(\mathbf{x}-\bm{\mu}_k)\right).
\end{equation}
and the complete parameter set is $\bm{\theta} := \{(\pi_k, \bm{\mu}_k, \bm{\Sigma}_k)\}_{k=1}^{K}$, with mixing weights satisfying
\begin{equation}
    \pi_k \geq 0 \quad \forall\, k, \qquad \sum_{k=1}^{K} \pi_k = 1.
\end{equation}
\\
The GMM admits an equivalent latent-variable formulation that can be utilized for clustering purposes. Introduce an unobserved categorical indicator $z \in \{1, \ldots, K\}$ with
\begin{equation}
    p(z = k) = \pi_k.
\end{equation}
Conditional on the cluster assignment $z = k$, the observation is drawn from the $k$-th Gaussian:
\begin{equation}
    p(\mathbf{x} \mid z = k) = \mathcal{N}(\mathbf{x} \mid \bm{\mu}_k, \bm{\Sigma}_k).
\end{equation}
\noindent
A GMM is fitted using the widely known Expectation Maximization (EM) algorithm, which is used for obtaining maximum likelihood estimates of parameters when some of the data is \textit{missing} or \textit{latent} \parencite{Dempster1977}. 
\\
In the present study, the observed variable $\mathbf{x} \in \mathbb{R}^5$ is a five-dimensional vector whose components correspond to a participant's self-reported ratings on five achievement emotions: joy, interest, hopelessness, boredom, and anxiety. The goal is to identify empirically grounded student profiles - that is, latent groupings of participants who exhibit similar emotional response patterns - in a data-driven manner, without imposing a priori assumptions about the number or composition of those profiles.
\\
After fitting the model, the cluster labels were hard assigned, given by the formulation: $\hat{z}_i = \operatorname*{arg\,max}_{k \in \{1,\ldots,K\}} p(z_i = k \mid \mathbf{x}_i, \bm{\theta})$ where $\mathbf{x}_i$ belongs to the cluster $\hat{z}_i = k$ given model parameters $\bm{\theta}$.

\subsection{Model Fitting and Selection}

The \texttt{mclust} package in \texttt{R} was used to fit candidate GMMs with different covariance structures and numbers of clusters. Model selection was based primarily on the Bayesian Information CIterion (BIC), while also considering cluster interpretability and balance.
\\
For the pre-lecture data, the three best-performing models were an ellipsoidal model with unequal volume and unequal orientation ($BIC = -1066.83, K = 5$, where $K$ denotes the number of clusters), an ellipsoidal model with unequal volume and common orientation ($BIC = -1078.21, K = 3$), and an ellipsoidal model with unequal volume and unequal orientation ($BIC = -1079.01, K = 4$). For the post-lecture data, the corresponding best-performing models were an ellipsoidal model with unequal volume and common orientation ($BIC = -1042.18, K = 3$), the same covariance structure with four clusters ($BIC = -1052.35, K = 4$), and an ellipsoidal model with unequal volume and unequal orientation ($BIC = -1053.78, K = 2$).
\\
Although the five-cluster model achieved the highest BIC for the pre-lecture data, the three-cluster solution was selected because it produced more balanced cluster sizes. The former resulted in two clusters with $n < 5$ without added interpretability.

\subsection{Cluster Visualization and Transition Analysis}
To visualize the five-dimensional emotional profiles, Principal Component Analysis (PCA) was used to project the data into two dimensions while preserving most of the variation. This aided interpretation of the identified clusters.
\\
The Pearson's correlation between clusters - identified before and after the lecture - showed a clear one-to-one correspondence and were labelled \emph{Enthusiastic}, \emph{Demotivated}, and \emph{Vulnerable} based on their emotional characteristics (see Results). Each participant's movement between clusters was then tracked from pre- to post-lecture. Transitions were classified as either \emph{intended} (movement toward a more favourable emotional state) or \emph{unintended}. A chi-squared test ($df=2$) was performed to examine whether instructor facial expression, students' gender and their grade at the time of testing influenced the type of transition.

\subsection{Exploratory Factor Analysis}
An exploratory factor analysis (EFA) was conducted to identify the latent psychological dimensions underlying the observed questionnaire variables related to confidence (general and content-specific), sense of utility, self-reported attention, motivation, engagement, social presence, and cognitive load. Since the variables were measured using ordinal Likert-type scales, a polychoric correlation matrix was computed to estimate the correlations between the underlying continuous latent responses. Sampling adequacy was evaluated using the Kaiser-Meyer-Olkin (KMO) measure, while Bartlett's test of sphericity was used to verify that the observed correlation matrix significantly deviated from an identity matrix. The obtained KMO and Bartlett statistics jointly indicated that the data were suitable for factor analysis.

Mathematically, the EFA assumes that the observed variables $\mathbf{x}$ arise from a smaller set of latent factors $\mathbf{f}$ according to
\[
\mathbf{x} = \mathbf{\Lambda}\mathbf{f} + \boldsymbol{\epsilon},
\]
where $\mathbf{\Lambda}$ denotes the factor loading matrix and $\boldsymbol{\epsilon}$ represents variable-specific residual variance. The corresponding covariance structure is expressed as
\[
\mathbf{\Sigma} = \mathbf{\Lambda}\mathbf{\Lambda}^\top + \mathbf{\Psi},
\]
where $\mathbf{\Psi}$ is a diagonal matrix containing uniqueness terms. The number of factors was determined using parallel analysis and scree-plot inspection (using \texttt{psych::fa.parallel} in \texttt{R}).

The EFA was implemented in \texttt{R} using the \texttt{psych} package. Factor extraction was performed using the minimum residual (\texttt{minres}) method, which estimates the loading matrix by minimizing the residual discrepancy between the observed and reproduced correlation matrices. Since the latent psychological constructs were expected to be correlated, an oblique \texttt{oblimin} rotation was applied to obtain a more interpretable factor structure. Factor scores were subsequently estimated using Bartlett's method for downstream visualization and modelling.

\subsection{Bayesian Path Analysis}
Say, $\{z_i\}$ denote the set of all questionnaire-based responses - centred and scaled. The path model is defined in three parts:
\\
Latent measurement model:
\\
$z_i = \lambda_i F_k + \epsilon_i$ where $k = 0 \text{ or } 1$
\\
$\lambda_i \sim \mathcal{N}(0,1), \epsilon_i \sim \mathcal{N}(0, \psi_i), \psi_i \sim \mathcal{\Gamma}(1, 0.5)$
\\
Regression models:
\\
$F_k = a_{k,0} + a_{k,1} I_\text{angry} + a_{k,2} I_\text{neutral} + a_{k,3} \text{ grade } + \zeta_k$ for $ k = 1, 2$
\\
$P(\text{switch} = 1) = \mathcal{L}(b_0 + b_1 I_\text{angry} + b_2 I_\text{neutral} + b_3 F_1 + b_4 F_2 + b_5 \text{ social-presence} + \varepsilon)$ where $a_{k,i} \sim \mathcal{N}(0,1), \varepsilon \sim \mathcal{N}(0,1), (\zeta_1, \zeta_2) \sim \mathcal{N}\Big(\mathbf{0}_2, 
\begin{pmatrix}
\phi_{11} & \phi_{12}\\
\phi_{12} & \phi_{22}
\end{pmatrix}\Big)
$

\section{Results}
\subsection{Preliminary Analyses}
\begin{figure}[!ht]
    \centering
    \includegraphics[width=.9\linewidth]{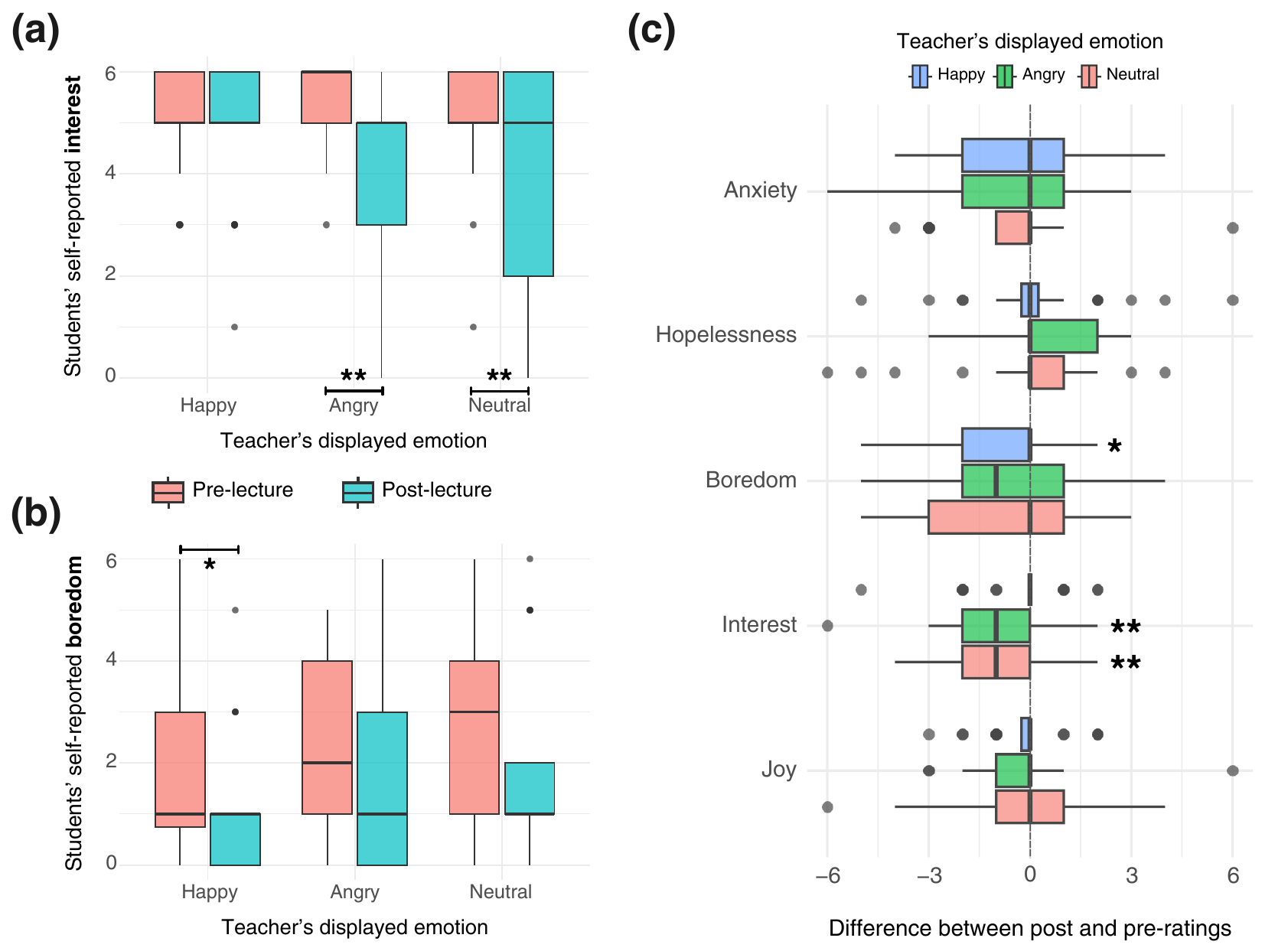}
    \caption{\textit{\textbf{(a)} Students' self-reported interest declines after the lecture with an angry or neutral instructor. \textbf{(b)} Self-reported boredom declines under a joyous instructor. \textbf{(c)} Change (post - pre) in self-reported achievement emotions. Although certain trends are observed (e.g., angry instructor increasing hopelessness and reducing joy), most are not statistically significant.}}
\end{figure}

To examine changes in students' achievement emotions from before to after the lesson, paired Wilcoxon signed-rank tests were conducted separately for each instructor condition. Among students taught by the positive instructor, only boredom showed a significant change from pre- to post-instruction ($V =$ 82, $p =$ .011), whereas joy, hope, interest, and anxiety remained stable (all $p >$ .05). In contrast, for both the neutral and negative instructor conditions, only interest changed significantly following instruction (neutral: $V =$ 185, $p =$ .002; negative: $V =$ 166, $p =$ .004), with no significant changes observed for joy, boredom, hope, or anxiety (all $p >$ .05). Overall, the results indicate that pre–post emotional changes were selective rather than widespread, affecting boredom only in the positive instructor condition and interest in the neutral and negative instructor conditions.
\\
Kruskal–Wallis tests were conducted to compare pre–post change scores across the three instructor conditions for each achievement emotion. A significant group difference was observed only for interest, $\chi^2(2) =$ 8.95, $p =$ .011, with larger declines in the angry and neutral instructor groups. No significant differences were found for changes in joy, boredom, hope, or anxiety (all $p >$ .05).

\begin{figure}[!ht]
    \centering
    \includegraphics[width=\linewidth]{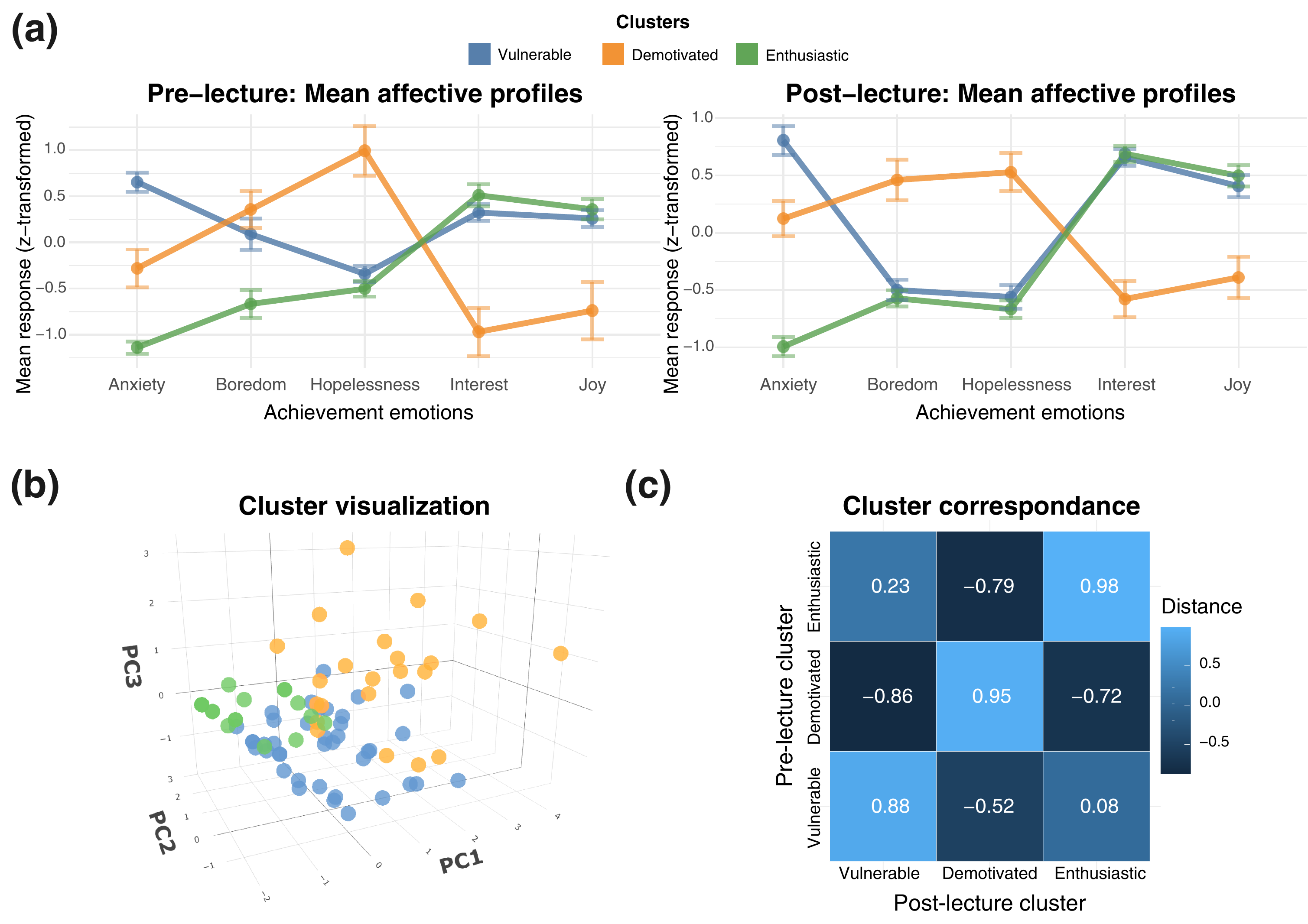}
    \caption{\textit{
    \textbf{(a)} Pre- and post-lecture achievement emotion profiles across five emotional dimensions. 
    \textbf{(b)} Projection of the profiles onto the first three principal components. 
    \textbf{(c)} Pearson's correlation establishing the correspondence between pre- and post-lecture profiles.
    }}
\end{figure}

\subsection{The Emotion Profiles: Interpreting the PCA Outcomes}
We first elaborate the Principal components which would enable us interpreting the GMM-based profiles in the next sub-section. We selected the first three components and the loadings are presented in the Table 2.
\\
\begin{table}[!ht]
    \centering
    \begin{tabular}{l c c c | c c c}
        \hline
        \textbf{Parameter} & \textbf{PC$_1$} & \textbf{PC$_2$} & \textbf{PC$_3$} & \textbf{PC$_1$} & \textbf{PC$_2$} & \textbf{PC$_3$}\\
        \hline
        \hline
         & \multicolumn{3}{l}{\textbf{ Pre-lecture}} & \multicolumn{3}{|l}{\textbf{Post-lecture}}\\
        \hline
        Variance explained (\%) & 29.6 & 21.2 & 18.6 & 28.6  & 21.2 & 18.2 \\
        Joy       & -0.515 & 0.158 & -0.310 & -0.400 & 0.456 &  0.352 \\
        Interest  & -0.509 & 0.347 & 0.224 & -0.462 & 0.169 & -0.800 \\
        Hopelessness &  0.458 & 0.129 & -0.720 &  0.517 & 0.245 & -0.474 \\
        Boredom   &  0.487 & 0.092 &  0.579 & 0.568 & -0.045 & -0.009 \\
        Anxiety   &  0.169 & 0.911 &  0.012 &  0.191 & 0.837 &  0.109 \\
        \hline
    \end{tabular}
    \caption{Principal component loadings (PC$_1$–PC$_3$) for pre- and post-lecture affect variables}
\end{table}
\noindent
The first principal component is primarily defined by loadings on joy, interest, hopelessness, and boredom. Notably, the positive emotions (joy and interest) exhibit loadings of similar sign and magnitude, whereas the negative emotions (hopelessness and boredom) load in the opposite direction. The second principal component is predominantly characterized by anxiety. Together, these patterns suggest that achievement emotions are not mutually independent, but instead exhibit a structured interdependence.

\subsection{Identifying and Interpreting the Emotion Profiles}
The Gaussian mixture model identified three profiles: ``Enthusiastic'' (high joy and interest, low hopelessness, boredom, and anxiety), ``Demotivated'' (high hopelessness and boredom, very low positive emotions), and ``Vulnerable'' (high joy and interest, low hopelessness and boredom, but elevated anxiety), shown in Figure 2 (a).
\\
Figure 2 (b) shows the demotivated profile clearly separated from the enthusiastic and vulnerable profiles along the PC$_1$ axis, consistent with PC$_1$ contrasting positive emotions with boredom and hopelessness. The PC$_2$ axis separates vulnerable from enthusiastic students, reflecting the dominant contribution of anxiety.
\\
Pearson's correlation coefficients between the mean affective profiles of the independently derived pre- and post-lecture clusters established their correspondence. As shown in Figure 2 (c), the strongest correlations occurred between qualitatively similar clusters, while the demotivated cluster showed consistently negative correlations with the enthusiastic and vulnerable clusters.

\subsection{Transition Analysis}
\begin{figure}[!ht]
    \centering
    \includegraphics[width=.9\linewidth]{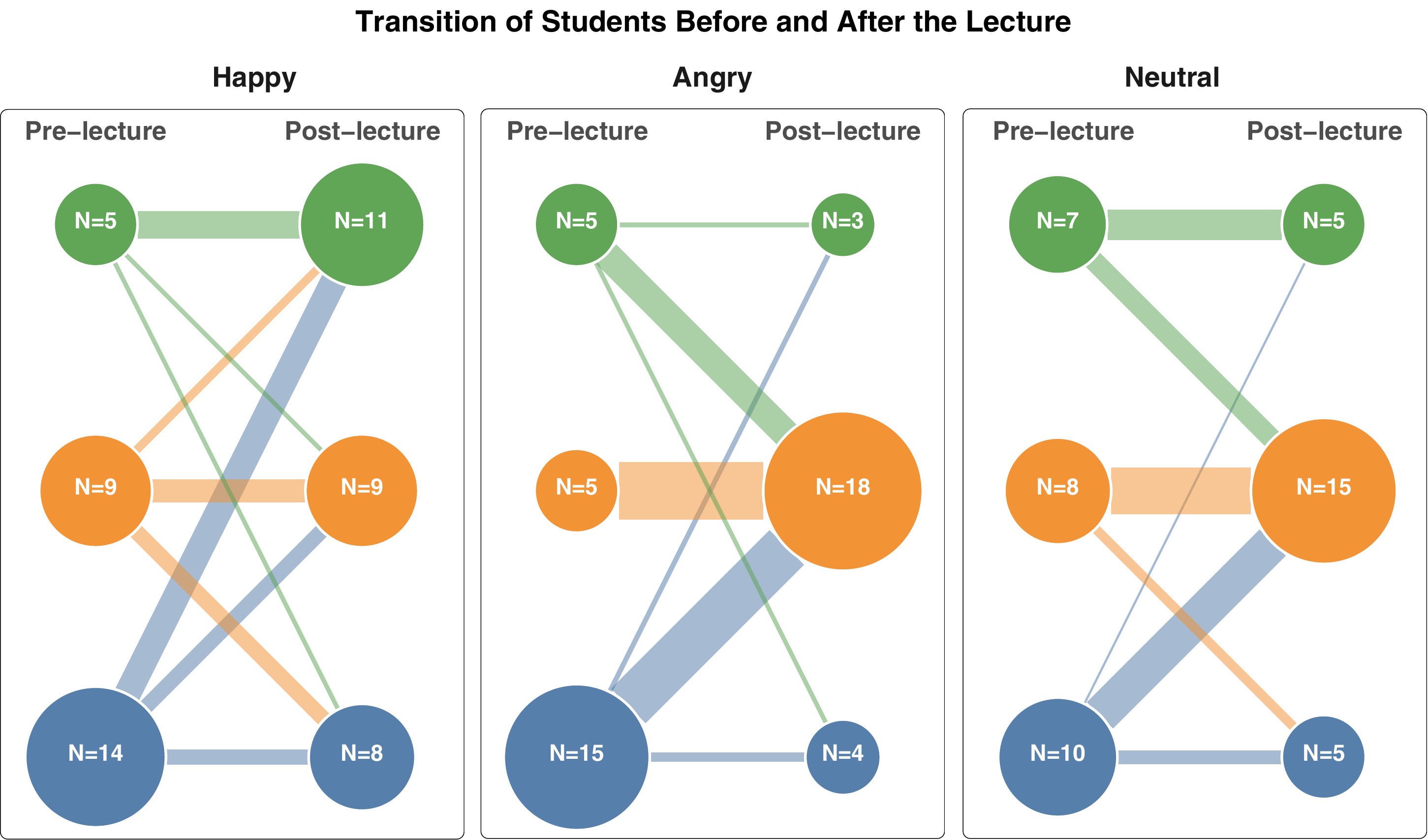}
    \caption{\textit{A visual representation of the learners transiting from their pre-lecture profile to the post-lecture one. The colour scheme is consistent with Figure 2.}}
\end{figure}
\noindent
We next examined transitions between the three profiles following the lecture. Figure 3 shows these transitions for the happy, angry, and neutral instructor conditions.
\\
To determine whether instructor facial expression influenced the transition type (`intended'' or `unintended'', defined in the Methodology), a Chi-squared test ($df = 2$) was conducted with the null hypothesis $H_0:$ teacher's emotion is independent of transition type. The null hypothesis was rejected ($\chi^2 = 9.08$, $p = 0.012$), indicating that transition type varied across instructor conditions. Intended transitions occurred most frequently in the happy condition (intended:unintended = $18:10$, $6:19$, and $10:15$ for the happy, angry, and neutral instructor conditions, respectively).

\begin{table}[!ht]
    \centering
    \begin{tabular}{c c c c @{\hspace{1.5cm}} c c @{\hspace{1.5cm}} c c}
    \hline
    & Happy & Angry & Neutral & Grade-VII & Grade-VIII & Girl & Boy \\
    \hline
    Intended   & 18 & 6  & 10 & 18 & 16 & 15 & 19 \\
    Unintended & 10 & 19 & 15 & 22 & 22 & 18 & 26 \\
    \hline
\end{tabular}
    \caption{Effect of instructor's emotion, student's grade and gender on transition: contingency table for the $\chi^2$ test}
\end{table}

\noindent
When the instructor displayed a happy emotion, the majority of students transitioned to the enthusiastic cluster. Under this condition, 55\% of the students initially classified as demotivated transitioned to either the vulnerable or the enthusiastic cluster. In the angry condition, however, most students transitioned to the demotivated cluster (67\% of those initially vulnerable and 60\% of those initially enthusiastic). Notably, all students who began in the demotivated cluster remained in that state. In the neutral condition, 60\% of the students initially in the vulnerable cluster and 75\% of those initially in the demotivated cluster were classified as demotivated at the end.
\\
However, we found no evidence of an association between profile transitions and either students’ grade ($df = 1$, $\chi^2 \approx 10^{-3}$) or gender ($df = 1$, $\chi^2 = 0.003$).

\subsection{Possible Cognitive Mechanism}
\begin{figure}[!ht]
    \centering
    \includegraphics[width=.65\linewidth]{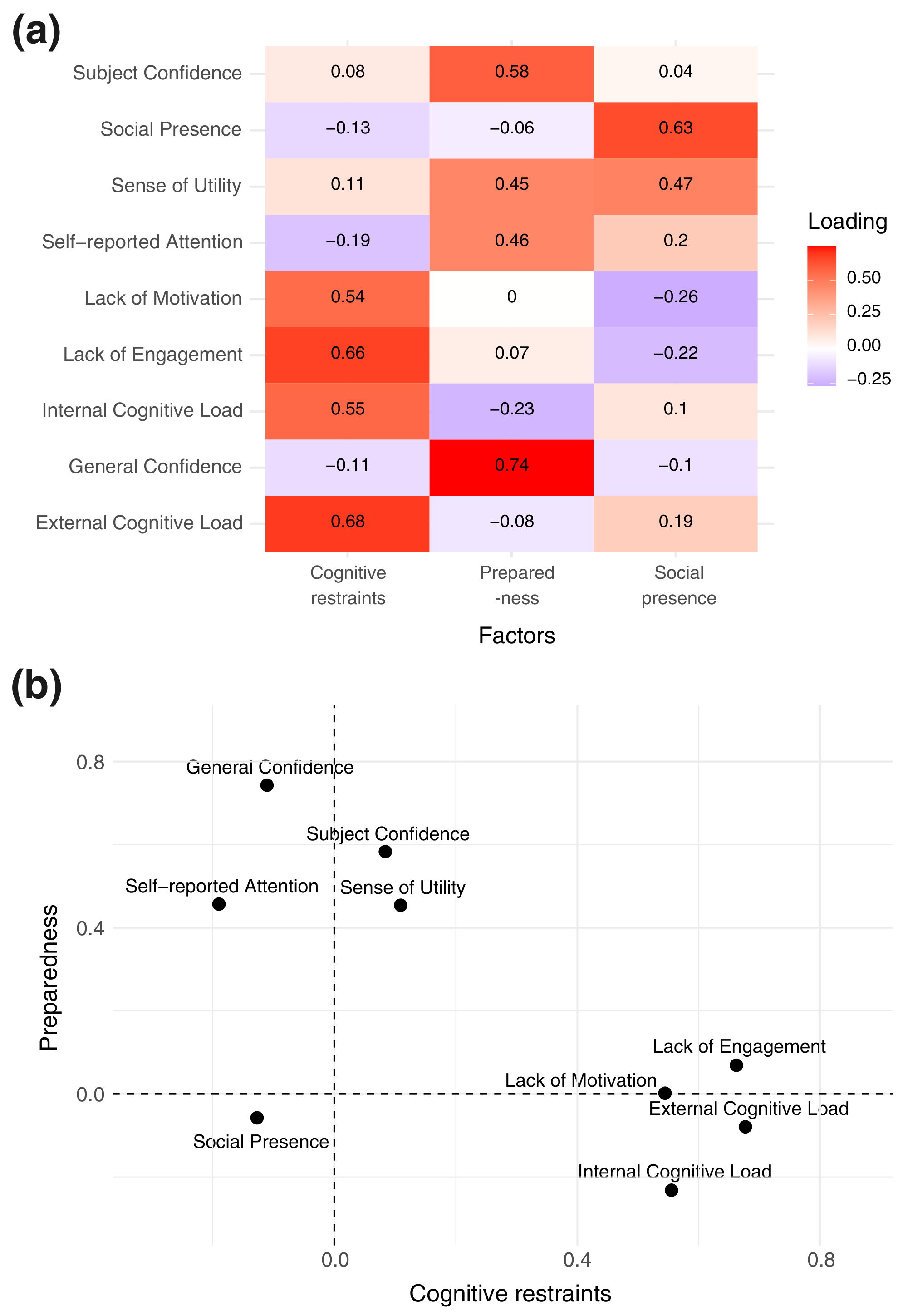}
    \caption{\textit{
        \textbf{(a)} Factor loadings of all target variables across the three factors identified through exploratory factor analysis.
        \textbf{(b)} A visualization of the orthogonal relationship between the two factors, ``cognitive restraint" and ``preparedness".
        }}
\end{figure}
In this section, we attempt to dissect the plausible mechanism through which instructors' facial expressions manifest in affective profile switching. We first conduct an exploratory factor analysis on students' responses to the preparedness-related questionnaire items (general confidence, topic-specific confidence, and perceived utility) and affective-cognitive state factors (motivation, engagement, perceived social presence, and internal and external cognitive load). We then utilize a Bayesian Structural Equation Modelling framework to examine and confirm the proposed mediation paths.

\subsubsection{Latent Regulatory Variables}
Before extracting the factors, the suitability of the polychoric correlation matrix for factor analysis was evaluated. The Kaiser-Meyer-Olkin measure of sampling adequacy was 0.71 overall, with item-level values ranging from 0.62 (social presence) to 0.81 (self-reported attention), indicating meritorious factorability. Bartlett's test of sphericity was significant, $\chi^2_{(df = 36)} = 151.90, p < .001$, confirming that the correlation matrix departed significantly from an identity matrix.
\\
Factor extraction was performed using minimum residual estimation (MINRES) with oblimin rotation, allowing factors to correlate freely. The number of factors to retain was determined by parallel analysis (1000 iterations), which indicated three factors.
\\
The solution demonstrated excellent fit to the polychoric correlation matrix. The likelihood ratio statistic was non-significant, $\chi^2_{(df = 12)}= 13.28, p = .35$, indicating no significant residual misfit. Supplementary fit indices confirmed adequacy: Tucker-Lewis Index = 0.966, RMSEA = 0.035, RMSR = 0.03, and BIC = -39. The three factors collectively accounted for 45\% of total item variance, with proportions of 19\%, 16\%, and 10\% for factors one through three respectively.
\\
The standardised pattern matrix is presented in Figure 4 (a). The first factor was defined by internal and external cognitive load, motivation, and engagement (due to a negative framing of the question). This factor was interpreted as cognitive restraint. The second factor was defined by general and subject-specific confidence, self-reported attention, and sense of utility, reflecting a preparedness and subject confidence dimension. The third factor was anchored by perceived social presence, with secondary contributions from the sense of utility ($\lambda = .47$) and self-reported attention ($\lambda = .20$); it was tentatively interpreted as perceived social and contextual relevance. However, the factor score adequacy for MR3 ($R^2 = .57$) was notably lower than for the first two factors ($R^2 = .67$ and $.66$, respectively), implying it might not be a stable latent construct in subsequent confirmatory analyses.
\\
Factor intercorrelations under the oblique solution were modest: $\rho(F1, F2) = -.34, \rho(F1, F3) = -.16$, and $\rho(F2, F3) = .13$, confirming that the three factors represent partially distinct psychological dimensions. 


\subsubsection{Confirmatory Analysis}
We tested alternative combinations of the EFA-derived indicators while retaining the four-item cognitive restraints factor (F1). The first model specified preparedness (F2) using general confidence, topic-specific confidence, perceived utility, and self-assessed attentiveness, with perceived social presence specified as a separate single-indicator factor (F3). This model showed acceptable fit, $\chi^2(25)=30.44$, $p=.208$. The second model combined social presence with the preparedness indicators to form a five-indicator F2. This model also showed acceptable fit, $\chi^2(26)=30.56$, $p=.245$, although the loading of social presence was not statistically significant ($p=.147$). The third model specified F2 using the same indicators as the second model, but excluding utility, while F3 comprised social presence and utility. Although this model yielded the best nominal fit, $\chi^2(24)=24.61$, $p=.427$, it produced a negative residual variance for utility, indicating a Heywood case and making the solution inadmissible. Consequently, the first model was retained as the most theoretically and statistically defensible measurement structure.

\subsubsection{Path Analysis}
\begin{figure}[!ht]
    \centering
    \includegraphics[width=\linewidth]{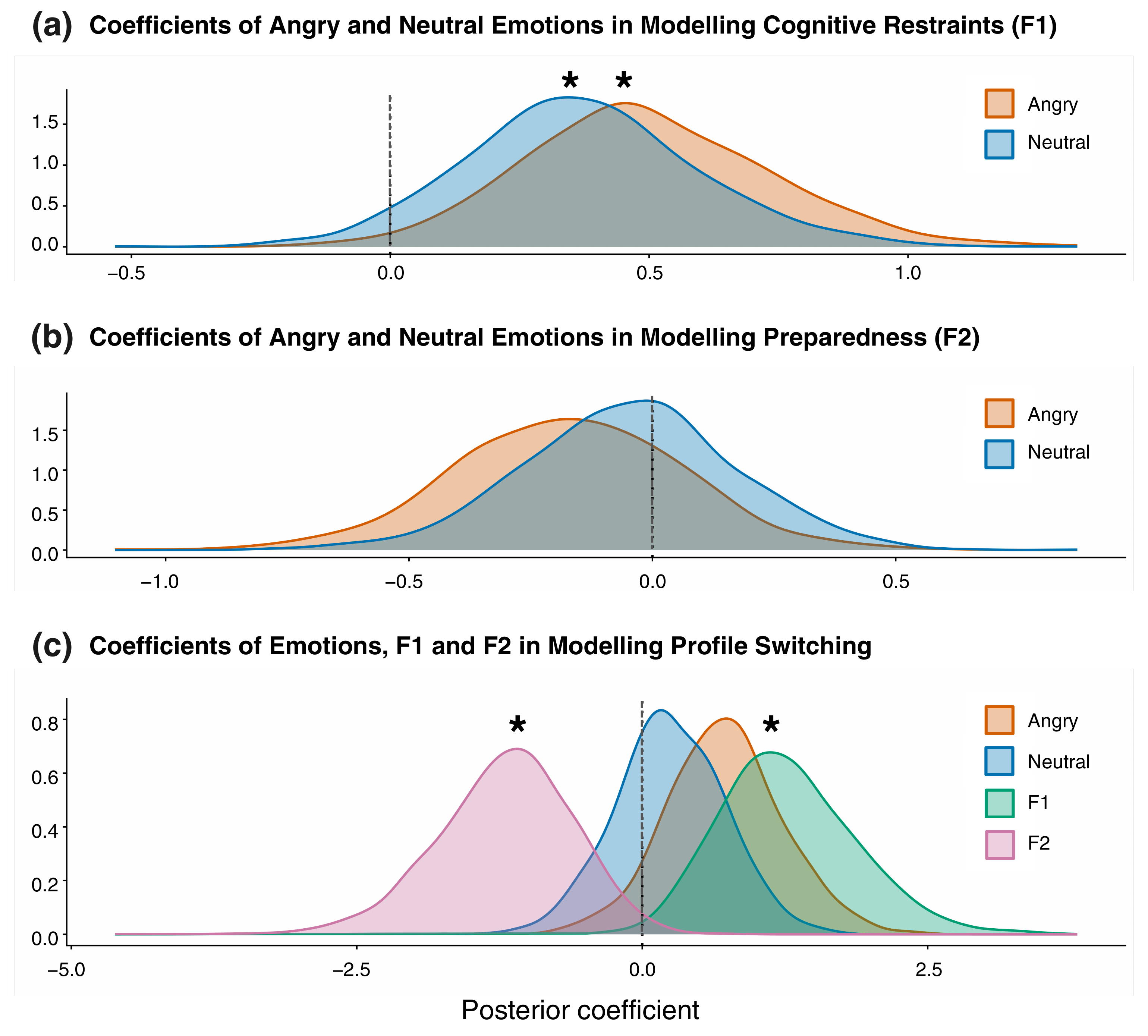}
    \caption{\textit{
    \textbf{(a)} In the Bayesian structural equation model, angry and neutral instructor emotions were associated with increased cognitive restraint (i.e., heightened cognitive load and reduced engagement and motivation), with happy emotion serving as the baseline.
    \textbf{(b)} Neither angry nor neutral emotion showed a clear association with the second latent factor, preparedness (confidence, perceived utility, and self-reported attention).
    \textbf{(c)} After accounting for the aforementioned pathways, neither emotion significantly predicted the probability of unintended switching, although angry emotion showed a positive tendency. In contrast, greater cognitive restraint was associated with more unintended switching, whereas greater preparedness was associated with less unintended switching; the 95\% credible intervals for both coefficients excluded zero.
    }}
\end{figure}

To examine the proposed mediation pathways, we fitted a Bayesian structural equation model (BSEM). The model showed adequate posterior predictive fit ($\mathrm{PPP}=0.254$), and the MCMC chains showed good convergence ($\hat{R}\leq1.005$ for all parameters). Instructor emotion was associated with F1, with the posterior mean for the Angry condition being positive ($\beta=0.496$, 95\% CI $[0.059,,0.981]$), whereas the corresponding effect of the Neutral condition was more uncertain ($\beta=0.367$, 95\% CI $[-0.057,,0.837]$). Neither emotion condition showed a clear association with F2, which is anticipated. 
\\
In the switch model, F1 showed a strong positive association with profile switching ($\beta=1.275$, 95\% CI $[0.255,,2.556]$), whereas F2 showed a strong negative association ($\beta=-1.203$, 95\% CI $[-2.467,,-0.156]$). The direct effects of instructor emotion on switching were comparatively uncertain. Accordingly, the posterior distribution of the total effect of Angry emotion on switching was positive ($\beta=1.558$, 95\% CI $[0.293,,2.823]$), whereas the corresponding total effect for Neutral emotion remained uncertain ($\beta=0.766$, 95\% CI $[-0.336,,1.869]$). 

\section{Discussion}
The present study examined the effect of an instructor's facial expression on learners' achievement emotions using a profile-based analytical framework. Conventional analyses revealed limited differences in individual emotions, i.e., pre–post changes in self-reported achievement emotions were limited to interest and boredom, with no significant differences across instructor conditions. However, the profile analysis identified systematic changes in learners’ emotional states and their transitions after experiencing specific facial expressions and intonations displayed by the instructor. Together, these findings suggest that modelling the learner as a whole, rather than analysing individual emotions independently, can provide a richer description of the teacher's displayed emotion.
\\
The profile-based approach identifies psychologically interpretable learner states that capture the coexistence of multiple emotions \parencite{Pekrun2006}. The vulnerable profile illustrates this distinction well: learners in this group reported high joy and interest alongside elevated anxiety, indicating that positive engagement and anxiety are not necessarily mutually exclusive. Such profiles provide a more realistic representation of learners' emotional experiences during instruction and support the earlier findings \parencite{Robinson2020_profscience, Radii2024_profprimary}.
\\
Another important aspect of the framework is the analysis of transitions between profiles. Classroom emotions are dynamic, making pre–post comparisons of individual emotions insufficient to capture their complexity. By linking pre- and post-lecture profiles, this approach tracks learners' transitions between psychologically meaningful affective states. Using this framework, we observed that instructor facial expression significantly influenced the nature of these transitions, with positive expressions being associated with more adaptive and intended transitions than angry or neutral expressions. Although changes in individual emotions in response to the instructor's expression were relatively inconclusive, the transition analysis revealed a coherent pattern at the profile level that would have been difficult to identify using univariate analyses alone. It is imperative to note that the comparatively weak effects observed in the individual emotion analyses should also be interpreted in the context of the experimental design and statistical power. Participants viewed a single eight-minute lecture delivered by an unfamiliar instructor in a controlled experimental setting. Under such conditions, large shifts in individual achievement emotions may not be expected.
\\
With the assertion that instructor expression shapes affective transitions, we next examined the psychological mechanisms that might explain this effect, using a structural model linking instructor condition, post-lecture psychological states, and the likelihood of an unintended transition.
\\
We first examined the structure of the post-lecture psychological measures using exploratory factor analysis. Despite relying on single-item self-reports, two psychologically coherent dimensions emerged. The first grouped cognitive load, motivation, and engagement together, and is best described as cognitive restraint: a state in which the learner feels burdened and disengaged. The second factor grouped confidence, perceived utility, and attention, reflecting preparedness: a state of readiness and positive appraisal of the subject. A third dimension, centred on social presence and perceived utility, was also suggested by the exploratory analysis, and is theoretically interesting, as it may reflect the learner's sense of connection to the instructor and the relevance of the content. However, this dimension proved unstable in confirmatory analyses, likely because the available indicators were too few and too ambiguous to define it cleanly at the current sample size. It is worth noting that self-reported attention loaded onto the preparedness factor rather than the cognitive restraint dimension. This is consistent with the idea that attention, as learners experience it, is less about cognitive effort and more about motivated, goal-directed engagement with the material.
\\
Confirmatory factor analysis supported the two-factor measurement model. The loadings were stable and in the theoretically expected directions. As mentioned earlier, the third factor did not replicate satisfactorily and was excluded from the structural model. Social presence was retained as a directly observed variable rather than a latent construct, given the absence of multiple indicators.
\\
Given the sample size of 78, reporting posterior distributions and credible intervals is more informative than relying on p-value thresholds, which may carry limited meaning when statistical power is constrained \parencite{Kruschke2017}. The structural model was therefore estimated using a Bayesian approach. The frequentist SEM, conducted separately, produced consistent results and serves as a robustness check.
\\
The structural findings suggest a coherent account of how instructor emotion shapes affective transitions. Angry and neutral instructor conditions both increased cognitive restraint relative to the joyful condition, i.e., students reported higher cognitive load, lower motivation, and lower engagement, with the angry condition producing the larger shift. This is consistent with existing work on emotional contagion and instructor affect \parencite{Zhang2023, Liu2021}. Instructor condition had no effect on preparedness, which is theoretically expected. A student's sense of confidence and readiness is likely shaped by prior knowledge and longer-term academic experience rather than a single eight-minute exposure to an instructor. The direction of the grade effect on preparedness was consistent with expectations: eighth-grade students tended toward higher preparedness than seventh-grade students. However, the effect was not credibly different from zero.
\\
The path from cognitive restraint to unintended switching was positive. Students who felt more burdened and disengaged were more likely to transition into an unintended affective state. The path from preparedness was negative and equally interpretable: students with a stronger sense of confidence and readiness were more likely to follow an intended trajectory. Critically, once the effect of instructor condition on restraint was accounted for, the direct path from instructor condition to switching was substantially reduced (Figure 5 (c)). This suggests that the influence of instructor emotion on affective transitions operates, at least in part, through its effect on the learner's cognitive and motivational state during the lecture. The total effect of angry instructor expression on unintended switching was robust across all model specifications (posterior estimate = 1.56, 95\% credible interval [0.29, 2.82]). The indirect effect through cognitive restraint was directionally consistent and had a posterior probability of approximately 0.93 of being positive, though its credible interval included zero: a result best interpreted as suggestive rather than conclusive at this sample size.
\\
The proposed pipeline opens several opportunities for future educational research. First, learner profiles provide an interpretable representation of classroom dynamics that extends beyond the traditional positive-negative dichotomy. Second, because profile correspondence can be established across repeated measurements, the same framework can be extended to longitudinal studies to examine how learners' emotional states evolve over longer instructional periods. Such analyses would be particularly valuable for comparing interactions with familiar and unfamiliar instructors, or for studying the development of learner-teacher relationships over time. Third, the third latent dimension suggested by the exploratory analysis (social presence and perceived utility) deserves direct investigation with properly designed multi-item measures. A study with sufficient power to establish this factor may find that the learner's sense of connection to the instructor and the perceived relevance of content constitute a distinct pathway through which instructor affect operates.
\\
\\
We acknowledge several limitations of the present study. Although the instructor portraying the happy, angry, and neutral conditions was a trained actor, no valence-arousal ratings were collected to verify whether the intended emotions were perceived as such. Moreover, the manipulation extended beyond facial expressions to include intonation and body language, making it difficult to attribute the observed transitions to facial expressions alone. We deliberately limited the lecture duration to under ten minutes to minimise fatigue, although such a brief intervention may not produce measurable changes in learners' affective states. The reliance on single-item measures for most psychological constructs is a further limitation: while the factor structure was recoverable, single items carry greater measurement error and limit the stability of latent constructs, as evidenced by the instability of the third factor. Finally, a sample size of 78 provides limited statistical power for clustering across five emotional dimensions, analysing transitions within a $3 \times 3 \times 3$ pre- to post-lecture profile matrix, and estimating a structural model with multiple latent variables. The mediation findings in particular should be treated as hypothesis-generating until confirmed in a larger sample.
\\
In summary, the present study demonstrates the utility of analysing achievement emotions as dynamic learner profiles rather than isolated emotional variables, and takes a step toward explaining why instructor expression shapes those profiles. The evidence points to cognitive restraint as a plausible mediating mechanism: angry instructor affect increases learner burden and disengagement, which in turn raises the likelihood of an unintended affective transition. Preparedness operates as an independent buffer. Although the sample size limits the strength of these conclusions, the framework and findings provide a concrete foundation for future work investigating the interplay between instructional behaviour, cognitive processes, and the evolution of learners' affective states in educational settings.

\printbibliography

@article{Pekrun2006,
  title = {The Control-Value Theory of Achievement Emotions: Assumptions,  Corollaries,  and Implications for Educational Research and Practice},
  volume = {18},
  ISSN = {1573-336X},
  url = {http://dx.doi.org/10.1007/s10648-006-9029-9},
  DOI = {10.1007/s10648-006-9029-9},
  number = {4},
  journal = {Educational Psychology Review},
  publisher = {Springer Science and Business Media LLC},
  author = {Pekrun,  Reinhard},
  year = {2006},
  month = nov,
  pages = {315–341}
}

@article{Pekrun2024,
  title = {Control-Value Theory: From Achievement Emotion to a General Theory of Human Emotions},
  volume = {36},
  ISSN = {1573-336X},
  url = {http://dx.doi.org/10.1007/s10648-024-09909-7},
  DOI = {10.1007/s10648-024-09909-7},
  number = {3},
  journal = {Educational Psychology Review},
  publisher = {Springer Science and Business Media LLC},
  author = {Pekrun,  Reinhard},
  year = {2024},
  month = aug 
}

@inbook{Pekrun2022_development,
  title = {Development of Achievement Emotions},
  ISBN = {9780191889516},
  url = {http://dx.doi.org/10.1093/oxfordhb/9780198855903.013.36},
  DOI = {10.1093/oxfordhb/9780198855903.013.36},
  booktitle = {The Oxford Handbook of Emotional Development},
  publisher = {Oxford University Press},
  author = {Pekrun,  Reinhard},
  year = {2022},
  month = jan,
  pages = {446–462}
}

@article{Qi2025_recent_meta,
  title = {The impact of achievement emotions on learning performance in online learning context: a meta-analysis},
  volume = {16},
  ISSN = {1664-1078},
  url = {http://dx.doi.org/10.3389/fpsyg.2025.1599543},
  DOI = {10.3389/fpsyg.2025.1599543},
  journal = {Frontiers in Psychology},
  publisher = {Frontiers Media SA},
  author = {Qi,  Mengmeng and Liu,  Wenying and Li,  Mingyue and Wang,  Gaoge and Liu,  Bowen},
  year = {2025},
  month = jul 
}

@article{CamachoMorles2021_recent_meta,
  title = {Activity Achievement Emotions and Academic Performance: A Meta-analysis},
  volume = {33},
  ISSN = {1573-336X},
  url = {http://dx.doi.org/10.1007/s10648-020-09585-3},
  DOI = {10.1007/s10648-020-09585-3},
  number = {3},
  journal = {Educational Psychology Review},
  publisher = {Springer Science and Business Media LLC},
  author = {Camacho-Morles,  Jesús and Slemp,  Gavin R. and Pekrun,  Reinhard and Loderer,  Kristina and Hou,  Hanchao and Oades,  Lindsay G.},
  year = {2021},
  month = jan,
  pages = {1051–1095}
}

@article{Iqbal2023,
  title = {Toward academic satisfaction and performance: the role of students’ achievement emotions},
  volume = {39},
  ISSN = {1878-5174},
  url = {http://dx.doi.org/10.1007/s10212-023-00751-z},
  DOI = {10.1007/s10212-023-00751-z},
  number = {3},
  journal = {European Journal of Psychology of Education},
  publisher = {Springer Science and Business Media LLC},
  author = {Iqbal,  Muhammad Zahid and Khan,  Tamania and Ikramullah,  Malik},
  year = {2023},
  month = oct,
  pages = {1913–1941}
}

@article{Kesevan2020,
  title = {Social Signalling as a Non Verbal Behaviour of Teachers in ESL Classroom Interaction},
  volume = {8},
  ISSN = {2332-3213},
  url = {http://dx.doi.org/10.13189/ujer.2020.081161},
  DOI = {10.13189/ujer.2020.081161},
  number = {11},
  journal = {Universal Journal of Educational Research},
  publisher = {Horizon Research Publishing Co.,  Ltd.},
  author = {Kesevan,  Hema Vanita and Madzlan,  Noor Alhusna and Hashim,  Haslinda},
  year = {2020},
  month = oct,
  pages = {5576–5580}
}

@article{Liu2021,
  title = {Does Teacher Immediacy Affect Students? A Systematic Review of the Association Between Teacher Verbal and Non-verbal Immediacy and Student Motivation},
  volume = {12},
  ISSN = {1664-1078},
  url = {http://dx.doi.org/10.3389/fpsyg.2021.713978},
  DOI = {10.3389/fpsyg.2021.713978},
  journal = {Frontiers in Psychology},
  publisher = {Frontiers Media SA},
  author = {Liu,  Wei},
  year = {2021},
  month = jun 
}

@article{Gu2024,
  title = {Onscreen presence of instructors in video lectures affects learners’ neural synchrony and visual attention during multimedia learning},
  volume = {121},
  ISSN = {1091-6490},
  url = {http://dx.doi.org/10.1073/pnas.2309054121},
  DOI = {10.1073/pnas.2309054121},
  number = {12},
  journal = {Proceedings of the National Academy of Sciences},
  publisher = {Proceedings of the National Academy of Sciences},
  author = {Gu,  Chanyuan and Peng,  Yingying and Nastase,  Samuel A. and Mayer,  Richard E. and Li,  Ping},
  year = {2024},
  month = mar 
}

@article{Wang2022important,
  title = {To Be Expressive or Not: The Role of Teachers’ Emotions in Students’ Learning},
  volume = {12},
  ISSN = {1664-1078},
  url = {http://dx.doi.org/10.3389/fpsyg.2021.737310},
  DOI = {10.3389/fpsyg.2021.737310},
  journal = {Frontiers in Psychology},
  publisher = {Frontiers Media SA},
  author = {Wang,  Yang},
  year = {2022},
  month = jan 
}

@article{Polat2022,
  title = {Instructors’ presence in instructional videos: A systematic review},
  volume = {28},
  ISSN = {1573-7608},
  url = {http://dx.doi.org/10.1007/s10639-022-11532-4},
  DOI = {10.1007/s10639-022-11532-4},
  number = {7},
  journal = {Education and Information Technologies},
  publisher = {Springer Science and Business Media LLC},
  author = {Polat,  Hamza},
  year = {2022},
  month = dec,
  pages = {8537–8569}
}

@article{Suen2024,
  title = {Enhancing learner affective engagement: The impact of instructor emotional expressions and vocal charisma in asynchronous video-based online learning},
  volume = {30},
  ISSN = {1573-7608},
  url = {http://dx.doi.org/10.1007/s10639-024-12956-w},
  DOI = {10.1007/s10639-024-12956-w},
  number = {3},
  journal = {Education and Information Technologies},
  publisher = {Springer Science and Business Media LLC},
  author = {Suen,  Hung-Yue and Hung,  Kuo-En},
  year = {2024},
  month = aug,
  pages = {4033–4060}
}

@article{Mega2014,
  title = {What makes a good student? How emotions,  self-regulated learning,  and motivation contribute to academic achievement.},
  volume = {106},
  ISSN = {0022-0663},
  url = {http://dx.doi.org/10.1037/A0033546},
  DOI = {10.1037/a0033546},
  number = {1},
  journal = {Journal of Educational Psychology},
  publisher = {American Psychological Association (APA)},
  author = {Mega,  Carolina and Ronconi,  Lucia and De Beni,  Rossana},
  year = {2014},
  month = feb,
  pages = {121–131}
}

@article{Wu2022,
  title = {Exploring the effects of achievement emotions on online learning outcomes: A systematic review},
  volume = {13},
  ISSN = {1664-1078},
  url = {http://dx.doi.org/10.3389/fpsyg.2022.977931},
  DOI = {10.3389/fpsyg.2022.977931},
  journal = {Frontiers in Psychology},
  publisher = {Frontiers Media SA},
  author = {Wu,  Rong and Yu,  Zhonggen},
  year = {2022},
  month = sep 
}

@article{Tze2020_profileuni,
  title = {Stability and change in the achievement emotion profiles of university students},
  volume = {41},
  ISSN = {1936-4733},
  url = {http://dx.doi.org/10.1007/s12144-020-01133-0},
  DOI = {10.1007/s12144-020-01133-0},
  number = {9},
  journal = {Current Psychology},
  publisher = {Springer Science and Business Media LLC},
  author = {Tze,  Virginia M. C. and Daniels,  Lia M. and Hamm,  Jeremy M. and Parker,  Patti C. and Perry,  Raymond P.},
  year = {2020},
  month = oct,
  pages = {6363–6374}
}

@article{Wang2021_profefl,
  title = {A latent profile analysis of EFL learners’ self-efficacy: Associations with academic emotions and language proficiency},
  volume = {103},
  ISSN = {0346-251X},
  url = {http://dx.doi.org/10.1016/j.system.2021.102633},
  DOI = {10.1016/j.system.2021.102633},
  journal = {System},
  publisher = {Elsevier BV},
  author = {Wang,  Yabing and Shen,  Bin and Yu,  Xiaoxiao},
  year = {2021},
  month = dec,
  pages = {102633}
}

@article{Chen2025_profmath,
  title = {A latent profile analysis of achievement emotions and their relationships with students’ perceived teaching quality in mathematics classrooms},
  volume = {155},
  ISSN = {0742-051X},
  url = {http://dx.doi.org/10.1016/j.tate.2024.104893},
  DOI = {10.1016/j.tate.2024.104893},
  journal = {Teaching and Teacher Education},
  publisher = {Elsevier BV},
  author = {Chen,  Xin and Zuo,  Haode and Lu,  Hong},
  year = {2025},
  month = mar,
  pages = {104893}
}

@article{Symes2025_profprimary,
  title = {Profiles of control,  value and achievement emotions in primary school mathematics lessons},
  volume = {95},
  ISSN = {2044-8279},
  url = {http://dx.doi.org/10.1111/bjep.12768},
  DOI = {10.1111/bjep.12768},
  number = {3},
  journal = {British Journal of Educational Psychology},
  publisher = {Wiley},
  author = {Symes,  Wendy and Lichtenfeld,  Stephanie and Wood,  Peter and Putwain,  David W.},
  year = {2025},
  month = mar,
  pages = {888–902}
}

@article{Radii2024_profprimary,
  title = {Scared,  Bored or Happy? Latent Profile Analyses of Primary School Students’ Academic Emotions about Math},
  volume = {14},
  ISSN = {2227-7102},
  url = {http://dx.doi.org/10.3390/educsci14080841},
  DOI = {10.3390/educsci14080841},
  number = {8},
  journal = {Education Sciences},
  publisher = {MDPI AG},
  author = {Radišić,  Jelena and Peixoto,  Francisco and Caetano,  Teresa and Mata,  Lourdes and Campos,  Mafalda and Krstić,  Ksenija},
  year = {2024},
  month = aug,
  pages = {841}
}

@article{Hickendorff2018,
  title = {Informative tools for characterizing individual differences in learning: Latent class,  latent profile,  and latent transition analysis},
  volume = {66},
  ISSN = {1041-6080},
  url = {http://dx.doi.org/10.1016/j.lindif.2017.11.001},
  DOI = {10.1016/j.lindif.2017.11.001},
  journal = {Learning and Individual Differences},
  publisher = {Elsevier BV},
  author = {Hickendorff,  Marian and Edelsbrunner,  Peter A. and McMullen,  Jake and Schneider,  Michael and Trezise,  Kelly},
  year = {2018},
  month = Aug,
  pages = {4–15}
}

@article{Robinson2020_profscience,
  title = {Momentary emotion profiles in high school science and their relations to control,  value,  achievement,  and science career intentions.},
  volume = {6},
  ISSN = {2333-8113},
  url = {http://dx.doi.org/10.1037/mot0000174},
  DOI = {10.1037/mot0000174},
  number = {4},
  journal = {Motivation Science},
  publisher = {American Psychological Association (APA)},
  author = {Robinson,  Kristy A. and Beymer,  Patrick N. and Ranellucci,  John and Schmidt,  Jennifer A.},
  year = {2020},
  month = dec,
  pages = {401–412}
}

@article{Robinson2017_profcollege,
  title = {Affective profiles and academic success in a college science course},
  volume = {51},
  ISSN = {0361-476X},
  url = {http://dx.doi.org/10.1016/j.cedpsych.2017.08.004},
  DOI = {10.1016/j.cedpsych.2017.08.004},
  journal = {Contemporary Educational Psychology},
  publisher = {Elsevier BV},
  author = {Robinson,  Kristy A. and Ranellucci,  John and Lee,  You-kyung and Wormington,  Stephanie V. and Roseth,  Cary J. and Linnenbrink-Garcia,  Lisa},
  year = {2017},
  month = oct,
  pages = {209–221}
}

@article{Jarrell2016,
  title = {The link between achievement emotions,  appraisals,  and task performance: pedagogical considerations for emotions in CBLEs},
  volume = {3},
  ISSN = {2197-9995},
  url = {http://dx.doi.org/10.1007/s40692-016-0064-3},
  DOI = {10.1007/s40692-016-0064-3},
  number = {3},
  journal = {Journal of Computers in Education},
  publisher = {Springer Science and Business Media LLC},
  author = {Jarrell,  Amanda and Harley,  Jason M. and Lajoie,  Susanne P.},
  year = {2016},
  month = jun,
  pages = {289–307}
}

@article{sachisthal2021trait,
  title={Trait and state math EAP (emotion, appraisals and performance) profiles of Dutch teenagers},
  author={Sachisthal, Maien SM and Raijmakers, Maartje EJ and Jansen, Brenda RJ},
  journal={Learning and Individual Differences},
  volume={89},
  pages={102029},
  year={2021},
  publisher={Elsevier}
}

@article{tempelaar2022types,
  title={Types of boredom and other learning activity emotions: A person-centred investigation of inter-individual data},
  author={Tempelaar, Dirk and Niculescu, Alexandra Corina},
  journal={Motivation and Emotion},
  volume={46},
  number={1},
  pages={84--99},
  year={2022},
  publisher={Springer}
}

@ARTICLE{Zhang2023,
  title     = "Teaching presence promotes learner affective engagement: The roles of cognitive load and need for cognition",
  author    = "Zhang, Yamei and Tian, Yuan and Yao, Liangshuang and Duan,
               Changying and Sun, Xiaojun and Niu, Gengfeng",
  journal   = "Teach. Teach. Educ.",
  publisher = "Elsevier BV",
  volume    =  129,
  number    =  104167,
  pages     = "104167",
  month     =  jul,
  year      =  2023,
  language  = "en"
}

@inproceedings{Sugiyo2018,
  title           = "The role of achievement emotions to cognitive load",
  booktitle       = "Proceedings of the International Conference on Science and
                     Education and Technology 2018 ({ISET} 2018)",
  author          = "Sugiyo, Mr and Sunawan, Mr and Pranoto, Yuli Kurniawati
                     Sugiyo",
  publisher       = "Atlantis Press",
  year            =  2018,
  address         = "Paris, France",
  conference      = "Proceedings of the International Conference on Science and
                     Education and Technology 2018 (ISET 2018)",
  location        = "Semarang, Indonesia"
}

@article{Sugiyo2021,
  title     = "Achievement goals and extraneous load predict germane load: The
               mediating effects of achievement emotions",
  author    = "Sugiyo Pranoto, Yuli Kurniawati",
  journal   = "Malays. J. Learn. Instr.",
  publisher = "UUM Press, Universiti Utara Malaysia",
  volume    =  18,
  number    =  2,
  pages     = "215--234",
  year      =  2021,
  language  = "en"
}

@article{sweller2019,
  title={Cognitive architecture and instructional design: 20 years later},
  author={Sweller, John and Van Merri{\"e}nboer, Jeroen JG and Paas, Fred},
  journal={Educational psychology review},
  volume={31},
  number={2},
  pages={261--292},
  year={2019},
  publisher={Springer}
}

@article{gunawardena1995,
  title={Social presence theory and implications for interaction and collaborative learning in computer conferences},
  author={Gunawardena, Charlotte N},
  journal={International journal of educational telecommunications},
  volume={1},
  number={2},
  pages={147--166},
  year={1995},
  publisher={Association for the Advancement of Computing in Education (AACE)}
}

@article{Bieleke2021,
  title = {The AEQ-S: A short version of the Achievement Emotions Questionnaire},
  volume = {65},
  ISSN = {0361-476X},
  url = {http://dx.doi.org/10.1016/j.cedpsych.2020.101940},
  DOI = {10.1016/j.cedpsych.2020.101940},
  journal = {Contemporary Educational Psychology},
  publisher = {Elsevier BV},
  author = {Bieleke,  Maik and Gogol,  Katarzyna and Goetz,  Thomas and Daniels,  Lia and Pekrun,  Reinhard},
  year = {2021},
  month = apr,
  pages = {101940}
}

@article{Dempster1977,
  title = {Maximum Likelihood from Incomplete Data Via the
                    <i>EM</i>
                    Algorithm},
  volume = {39},
  ISSN = {1467-9868},
  url = {http://dx.doi.org/10.1111/j.2517-6161.1977.tb01600.x},
  DOI = {10.1111/j.2517-6161.1977.tb01600.x},
  number = {1},
  journal = {Journal of the Royal Statistical Society Series B: Statistical Methodology},
  publisher = {Oxford University Press (OUP)},
  author = {Dempster,  A. P. and Laird,  N. M. and Rubin,  D. B.},
  year = {1977},
  month = Sept,
  pages = {1–22}
}

@article{Kruschke2017,
  title = {The Bayesian New Statistics: Hypothesis testing,  estimation,  meta-analysis,  and power analysis from a Bayesian perspective},
  volume = {25},
  ISSN = {1531-5320},
  url = {http://dx.doi.org/10.3758/s13423-016-1221-4},
  DOI = {10.3758/s13423-016-1221-4},
  number = {1},
  journal = {Psychonomic Bulletin \&amp; Review},
  publisher = {Springer Science and Business Media LLC},
  author = {Kruschke,  John K. and Liddell,  Torrin M.},
  year = {2017},
  month = Feb,
  pages = {178–206}
}

@article{Marici2025,
  title = {The role of teachers’ emotional facial expressions on student perceptions and engagement for primary school students-an experimental investigation},
  volume = {16},
  ISSN = {1664-1078},
  url = {http://dx.doi.org/10.3389/fpsyg.2025.1613073},
  DOI = {10.3389/fpsyg.2025.1613073},
  journal = {Frontiers in Psychology},
  publisher = {Frontiers Media SA},
  author = {Marici,  Marius and Iosim,  Iasmina and Marin,  Cornelia Diana},
  year = {2025},
  month = Aug 
}
\end{document}